\documentclass[numberedappendix,apj,twocolumn]{emulateapj}

\usepackage[bookmarks,bookmarksnumbered,colorlinks=true, citecolor=blue, linkcolor=black,breaklinks]{hyperref}

\usepackage{color}

\newcommand\Tstrut{\rule{0pt}{2.6ex}}         % = `top' strut
\newcommand\Bstrut{\rule[-0.9ex]{0pt}{0pt}}   % = `bottom' strut

\shorttitle{$z\sim10$ UV LF from All Legacy HST Fields}
\shortauthors{Oesch et al.}

\submitted{Draft version \today}

\begin{document}

\title{The Dearth of $z\sim10$ Galaxies in all HST Legacy Fields -- \\ The Rapid Evolution of the Galaxy Population in the First 500 Myr
\altaffilmark{1}}

\altaffiltext{1}{Based on data obtained with the \textit{Hubble Space Telescope} operated by AURA, Inc. for NASA under contract NAS5-26555. }

\author{P. A. Oesch\altaffilmark{2},
R. J. Bouwens\altaffilmark{3},
G. D. Illingworth\altaffilmark{4}, 
I. Labb\'{e}\altaffilmark{3},
M. Stefanon\altaffilmark{3}
}

\altaffiltext{2}{Geneva Observatory, University of Geneva, Ch. des Maillettes 51, 1290 Versoix, Switzerland; \email{pascal.oesch@unige.ch}}
\altaffiltext{3}{Leiden Observatory, Leiden University, NL-2300 RA Leiden, Netherlands}
\altaffiltext{4}{UCO/Lick Observatory, University of California, Santa Cruz, CA 95064, USA}

\begin{abstract}
We present an analysis of all prime $HST$ legacy fields spanning $>800$ arcmin$^2$ for the search of $z\sim10$ galaxy candidates and the study of their UV luminosity function. In particular, we present new $z\sim10$ candidates selected from the full Hubble Frontier Field (HFF) dataset. Despite the addition of these new fields, we find a low abundance of $z\sim10$ candidates with only 9 reliable sources identified in all prime HST datasets that include the HUDF09/12, the HUDF/XDF, all the CANDELS fields, and now the HFF survey. Based on this comprehensive search, we find that the UV luminosity function decreases by one order of magnitude from $z\sim8$ to $z\sim10$ over a four magnitude range. This also implies a decrease of the cosmic star-formation rate density by an order of magnitude within 170 Myr from $z\sim8$ to $z\sim10$. We show that this accelerated evolution compared to lower redshift can entirely be explained by the fast build-up of the dark matter halo mass function at $z>8$. Consequently, the predicted UV LFs from several models of galaxy formation are in good agreement with this observed trend, even though the measured UV LF lies at the low end of model predictions. The difference is generally still consistent within the Poisson and cosmic variance uncertainties. We discuss the implications of these results in light of the upcoming James Webb Space Telescope mission, which is poised to find much larger samples of $z\sim10$ galaxies as well as their progenitors at less than 400 Myr after the Big Bang.
\vspace*{0.2truecm}
\end{abstract}


\vspace*{0.4truecm}

\section{Introduction}

Understanding the formation and evolution of the first generations of galaxies in the early universe is still one of the most challenging and intriguing questions of modern observational astronomy. Thanks to the availability of sensitive near-infrared data taken with the {\it Hubble Space Telescope's} ($HST$) Wide Field Camera 3 (WFC3) over the last few years, the exploration of galaxies has now reached $z\sim10-12$, less than 500 Myr after the Big Bang \citep[e.g.][]{Bouwens11a,Bouwens16z910,Ellis13,Coe13,Oesch14,Oesch16,McLeod16,Ishigaki17}.

In particular, out to $z\sim8$, large galaxy samples have now been identified and used for the study of galaxy build up \citep[e.g.][]{Bouwens11c,Bouwens15UVLFs,Bradley13, Finkelstein12b,Schenker13,Mclure13,Schmidt14,BaroneNugent14,Stefanon17b}. These galaxy samples enabled accurate measurements of the UV LF and the SFRD, from which a consensus picture emerged. Between $z\sim8$ and $z\sim3$, there is general agreement that galaxies build up at a remarkably steady rate of about a factor $2\times$ growth per redshift bin \citep[see e.g.][for recent reviews]{Stark16,Finkelstein16}. 

At even higher redshift, $z>8$, the situation becomes less clear, mainly due to small galaxy samples in previous datasets. 
While the analysis of the full HUDF09/12 and CANDELS GOODS data revealed a rapid, accelerated evolution of the SFRD by $\sim10\times$ from $z\sim8$ to $z\sim10$ in only 170 Myr \citep[see e.g.][but see also Ellis et al. 2013\nocite{Ellis13}]{Oesch12a,Oesch14}, the two detections of $z>9$ galaxies in the small volume probed by the CLASH survey \citep{Zheng12, Coe13} were consistent with less evolution from $z\sim8$ to $z\sim10$ \citep[see also][]{McLeod16}. 
However, these early results had large uncertainties since they were mostly based on a few individual sources identified in small survey volumes.

Apart from the astrophysical implication of these different results on the $z>8$ SFRD, understanding the evolution of the galaxy number counts to $z>8$ is particularly important in preparation for the next milestone in extragalactic astronomy, the launch of the \textit{James Webb Space Telescope} ($JWST$). Given the limited lifetime of \textit{JWST}, it is crucial to obtain reliable predictions of how the UV LF evolves to $z>8$ in order to prepare the most efficient surveys and maximally exploit the telescope.

The HFF program \citep{Lotz16} is ideally suited to provide new constraints by providing additional search volume and larger samples of galaxies at $z\sim10$. The HFF exploits the lensing magnification of six massive foreground clusters in order to probe intrinsically very faint sources, fainter than accessible with the deepest $HST$ data over the HUDF, but over a reduced volume. Additionally, the HFF observes six deep parallel blank field pointings, which help to mitigate the uncertainties of magnification maps and cosmic variance \citep[see][]{Coe14,Lotz16}.

Several authors have already exploited the HFF dataset for high-redshift searches extending out to $z\sim10$ \citep[][]{Zitrin14,Oesch15,Infante15,Ishigaki15,Ishigaki17,McLeod16}. However, most HFF analyses so far have studied the HFF images separately from previous HST datasets, and often only reported results based on an early subset of the final HFF data. The main goal of this paper is to finally exploit \textit{all} the legacy $HST$ datasets {\it together}, including the previous fields and the full HFF data, and to analyze these in a consistent manner to reach the best-possible constraints on the evolution of the galaxy population at $z>8$ and on the UV LF at $z\sim10$ before the advent of $JWST$. In particular, a major goal of the present paper is to test the accelerated evolution of the galaxy population and the SFRD at $z>8$ that is still debated in the literature.

Specifically, this paper is organized as follows. In Section 2, we outline the full $HST$ dataset that is used here, which includes all extragalactic $HST$ legacy imaging fields. In Section 3, we present our selection which lead to new $z\sim10$ galaxy candidates identified in the latest HFF data, and we also list the candidates from previous datasets.
A small sample of additional, possible $z\sim10$ candidates from the HFF dataset that still need deeper data to be confirmed is listed in the appendix.
The resulting UV LFs and SFRD estimates from the combined data are shown in Section 4, before we close with a summary (Section 5).

Throughout this paper, we adopt $\Omega_M=0.3, \Omega_\Lambda=0.7, H_0=70$ kms$^{-1}$Mpc$^{-1}$, i.e. $h=0.7$, consistent with the most recent measurements from Planck \citep{Planck2015}. Magnitudes are given in the AB system \citep{Oke83}, and we will refer to the HST filters F435W, F606W, F814W, F105W, F125W, F140W, F160W as $B_{435}$, $V_{606}$, $I_{814}$, $Y_{105}$, $J_{125}$, $JH_{140}$, $H_{160}$, respectively.

\begin{deluxetable}{lccc}
\tablecaption{Fields included in $z\sim10$ search in this paper\label{tab:fields}}
\tablewidth{\linewidth}
\tablecolumns{4}
\tablehead{Field & Area [arcmin$^2$]  & Depth\tablenotemark{$\dagger$}  & Ref.\tablenotemark{*} }

\startdata

HUDF12/XDF  &  4.7   & 29.8  & 1 \\  
HUDF09-1    &  4.7   & 29.0  & 1 \\
HUDF09-2    &  4.7   & 29.3  & 1 \\
ERS         &  41.3   & 28.0  & 1 \\
GOODSS-Deep   &  63.1   & 28.3  & 1 \\
GOODSS-Wide  &  41.9   & 27.5  & 1 \\

GOODSN-Deep &  64.5  & 27.8  & 2 \\ 
GOODSN-Wide &  69.4  & 27.1   & 2\Bstrut\\

\hline

CANDELS/EGS \Tstrut  & 170  & 26.6  & 3  \\
CANDELS/UDS  & 150  & 26.5  & 3 \\
CANDELS/COSMOS  & 150  & 26.3   & 3\Bstrut\\

\hline
HFF (6 cluster + 6 parallel)   &  56.4  & 28.7   & \Tstrut \Bstrut \\

\hline

\textbf{Total HST}  & \textbf{821} & 26.3-29.8    &  \Bstrut \rule{0pt}{3.5ex}

\enddata

\tablenotetext{$\dagger$}{5$\sigma$ depth in AB magnitudes, measured in apertures of 0\farcs35 diameter}
\tablenotetext{*}{Previous $z\sim10$ searches by our team included in this analysis. 1: \citet{Oesch12a}, 2: \citet{Oesch14}, 3: \citet{Bouwens16z910}}

\end{deluxetable}

\section{Data Set}
\label{sec:data}

\subsection{Ancillary Legacy HST Dataset}

In this paper, we combine all legacy $HST$ datasets that have deep optical and NIR imaging for a search of $z\sim10$ galaxy candidates. In particular, we include all the imaging data that have been analyzed by our team in \citet{Oesch12a}, \citet{Oesch14}, and \citet{Bouwens16z910}. This includes the deepest WFC3/IR and ACS data available over the Hubble Ultra Deep Field (HUDF) / eXtreme Deep Field \citep[XDF][]{Illingworth13,Ellis13}, the deep parallel fields from the UDF05/HUDF09 surveys \citep[][]{Oesch07,Bouwens11c}, the WFC3 Early Release Science (ERS) images \citep[][]{Windhorst11}, as well as the imaging from all the five fields of the CANDELS survey \citep{Grogin11,Koekemoer11}. The $5\sigma$ depths in these images range from $H_{AB}=26.3$ mag over the CANDELS Wide fields to $H_{AB} = 29.8$ mag over the small area in the XDF. Most importantly, all these fields are covered by WFC3 $J_{125}$ and $H_{160}$ imaging as well as shorter wavelength $HST$ data that we require for the selection of $z\sim10$ galaxy candidates.  A full list of all fields included in the analysis as well as their areas and depths can be found in Table \ref{tab:fields}. For a detailed description of these datasets we refer the reader to our previous papers referenced in the table.

\subsection{Hubble Frontier Field Dataset}

The latest $HST$ dataset comes from the HFF program, which obtained very deep images over six clusters and six parallel fields for 140 orbits each, split over seven filters \citep[see][]{Lotz16}. We have searched for $z\sim10$ galaxy candidates in the first of these cluster/parallel fields \citep[A2744;][see also Zitrin et al. 2014, McLeod et al.\ 2016, Ishigaki et al.\ 2017]{Oesch15}\nocite{Zitrin14}. Here, we now extend our analysis to the completed HFF dataset, which includes 12 WFC3/IR fields. In particular, we use the fully reduced version 1 images provided by STScI of all HFF fields at a pixel scale of 60 mas\footnote{\url{http://archive.stsci.edu/pub/hlsp/frontier/}}. These images have a $5\sigma$ depth of $H_{160}=28.7$ mag as measured in circular apertures of 0\farcs35 diameter in empty sky regions. 

In order to minimize the impact of intra-cluster light (ICL) as well as the outskirts of very bright and extended cluster galaxies, we subtract a 2\farcs5 wide median filtered image of all HFF cluster data. The cores of bright sources are excluded in the filtering process which minimizes over-subtraction around bright galaxies or stars. This procedure allows us to select faint galaxies well into the cluster core \citep[see also][]{Oesch15}. Several authors have developed comparable techniques to deal with the ICL \citep[e.g.][]{Atek15,Merlin16,Bouwens17HFF1,Livermore17}. While the goal of all these procedures is to obtain as complete a high-redshift galaxy sample as possible, the exact procedure used for the galaxy search is not necessarily that important, as long as the detection completeness of the resulting ICL-subtracted dataset is properly quantified through adequate simulations (see Section \ref{sec:completeness}).

%\subsection{Lens Models}

To account for gravitational lensing by the foreground clusters in the HFF dataset, we exploit the public lens models made available by several teams on the MAST Frontier Field webpage\footnote{\url{archive.stsci.edu/prepds/frontier/lensmodels/}}. For the last two clusters that were observed by the HFF campaign (As1063, A370) these models are only based on multiple images identified in ancillary data taken \textit{before} the HFF campaign. However, for the first four clusters (A2744, MACS0416, MACS0717, and MACS1149), we use updated models (v3) that are based on a much larger number of multiple images that have been found in the HFF dataset. In particular, for those four clusters, our baseline lens model throughout this paper will be based on the glafic code \citep{Oguri10} as described in \citet{Kawamata16}. For the last two clusters, we base our analysis on the models by Zitrin et al \citep[e.g.,][]{ZitrinModel}, who also released both components of the shear tensor allowing us to compute the radial and tangential magnification factors to properly estimate the selection volume of high redshift galaxies as discussed in \citet{Oesch15}. We have tested and verified that our results do not change significantly, when using different lens models. 

A small amount of magnification is also present in the parallel fields, which is estimated in the models of \citet{MertenModel} using weak lensing. The typical magnification is of order 10-15\%, which we account for as well.

\subsection{Spitzer/IRAC Dataset}

Longer wavelength constraints from deep Spitzer/IRAC images are extremely important for the search of very high-redshift galaxies due to potential contamination by lower redshift interlopers \citep[see e.g.][]{Oesch12a,Holwerda15,Vulcani17}. In particular, dusty or quiescent galaxies can exhibit similarly red colors and remain undetected at shorter wavelengths, which are the main features used to select high-redshift galaxies. This problem is exacerbated for $z\sim10$ galaxy searches, as the most distant sources are only detected in the longest wavelength $HST$ filter ($H_{160}$). 

To mitigate this problem, we analyze deep Spitzer/IRAC data at 3.6~$\mu$m and 4.5~$\mu$m that are available over all the HST fields used here. In particular, ultra-deep data IRAC data over the HUDF and the two GOODS fields are available as part of a number of programs, including the IUDF, iGOODS, and GREATS \citep[][Labbe et al. 2017, in prep.]{Labbe15}. The remaining CANDELS fields have been covered by the S-CANDELS program \citep{Ashby15}, and deep Spitzer/IRAC data over the HFFs have been obtained as part of a director's discretionary time program\footnote{A list of all the Spitzer programs covering the HFF fields can be found here: \url{http://irsa.ipac.caltech.edu/data/SPITZER/Frontier/}}.

We use our own Spitzer/IRAC reductions that were produced using our well-tested pipeline including all the data in the IRSA archive over these fields. The images were aligned to the HST $H_{160}$ data and were drizzled to a pixel scale of 120 mas (i.e., twice the pixel scale of the HST images). For more information on the reduction pipeline see \citet{Labbe15}. The depths of the IRAC images varies significantly, but reaches as faint as 27.2 mag (3$\sigma$ at 3.6~$\mu$m) in the deepest regions in the GOODS-South and -North fields with an exposure time of up to 200 hr thanks to the latest IRAC data from the GREATS survey (Labbe et al. 2017, in prep).

\setlength{\tabcolsep}{0.04in} 
\begin{deluxetable*}{lcccccccc}
\tablecaption{Photometry of $z\sim9-10$ Candidates in all the 6 HFF Clusters \label{tab:phot}}
\tablewidth{0.86\linewidth}
\tablecolumns{9}
%\tabletypesize{\tiny}

\tablehead{\colhead{ID} & R.A. & Decl. &\colhead{$H_{160}$}  & \colhead{$J_{125}-H_{160}$}  & \colhead{S/N$_{160}$}    & \colhead{$\mu$\tablenotemark{*}} & \colhead{$M_{UV}$\tablenotemark{\dag}} & \colhead{Reference}}

\startdata

\cutinhead{Abell 2744}
A2744-JD1A  &  00:14:22.20  &  -30:24:05.3  &  $26.83\pm0.15$ & $1.2 \pm 0.2 $   & 14.0  & 13.8 ($8-23$)  \Tstrut &  $-17.8\pm0.2$ & 1,2 \\  % 5684 %, 12.9
A2744-JD1B  &  00:14:22.80  &  -30:24:02.8  &  $26.34\pm0.09$ & $1.4 \pm 0.2 $  & 14.3   & 26.4 ($7-27$) & $-17.6\pm0.6$  & 1,2 \\  % 5887, , 11.2

\cutinhead{Abell 370}

A370par-JD1  &  02:40:10.62  &  -01:37:31.2  &  $27.85\pm0.16$ & $1.4 \pm 0.5 $   & 10.1  & $1.16^{+0.09}_{-0.07}$  & $-19.5\pm0.2$ & -- \\  
A370par-JD2  &  02:40:14.92  &  -01:38:04.3  &  $27.90\pm0.23$ & $> 1.7$   & 6.4  & $1.10^{+0.08}_{-0.06}$ & $-19.5\pm0.3$ & --   % 848

\enddata

\tablerefs{(1) \citet{Zitrin14}, (2) \citet{Oesch15}  }
\tablenotetext{*}{Magnification numbers quoted for the clusters are derived from the Glafic (v3) magnification maps, while the numbers in the brackets show the range of magnifications from other models. For the candidates in the parallel field, the magnification number with errorbars come from the Merten v1 model. } %and the CATS v4.1 
\tablenotetext{$\dagger$}{Absolute magnitudes are based on the Glafic lensing models, but include a systematic errorbar that accounts for the model uncertainties. }

\end{deluxetable*}

\section{Galaxy Sample}
\label{sec:sample}

\subsection{The $z\sim10$ Lyman Break Selection}

Our basic sample selection is the same in all $HST$ fields. It is derived from a catalog based on a $\chi^2$ detection image constructed from the $H_{160}$ and $JH_{140}$ images, when the latter is available. For fields without $JH_{140}$ data, the detections are simply based on the $H_{160}$ images. Source photometry is measured with SExtractor \citep{Bertin96} run in dual image mode.  
All images are downgraded to the $H_{160}$-band point-spread function through the convolution with an appropriate kernel derived from stars in the fields. 

Galactic extinction is accounted for by adjusting the zeropoints for each HST filter using a Milky Way extinction curve \citep{Cardelli89} and $E(B-V)$ values based on the maps of \citet{Schlafly11}\footnote{\url{http://irsa.ipac.caltech.edu/applications/DUST/}}. Corrections are typically quite small with $<0.01$ mag in the WFC3/IR filters and $<0.1$ mag in the $B_{435}$ filter. However, in one case, the corrections reach up to 0.3 mag in $B_{435}$ for the MACS0717 field, which has a Galactic extinction of $E(B-V)=0.066$ mag.

Galaxies at $z > 9.5$ are identified by exploiting the spectral break at the Ly$\alpha$ line due to almost complete absorption by neutral inter-galactic hydrogen.  At this redshift, the Ly$\alpha$ break shifts into the $J_{125}$ filter, thus resulting in a red $J_{125}-H_{160}$ color and a non-detection in shorter wavelength filters. 
Following previous analyses by our team \citep[e.g.][]{Oesch14}, we restrict the search here to galaxies with $J_{125}-H_{160}>1.2$, which selects sources at $z\gtrsim9.5$. 

In particular, our color selection and non-detection criteria are:
\begin{equation}
	(J_{125}-H_{160})>1.2
\end{equation}
\[
S/N(B_{435} \mathrm{ ~to~ } Y_{105})<2\quad 
\]

Only sources with a $H_{160}>5\sigma$ are considered. Where $JH_{140}$ data are available, we also consider sources with $>3.5\sigma$ detections in each $H_{160}$ and $JH_{140}$ and at least $>5\sigma$ in one of the bands. In the HFF fields, we used a slightly higher signal-to-noise cut of S/N$>6$ in the $H_{160}$ band to limit the impact of correlated noise and residuals in the ICL background subtraction (see also next section).

In addition to the non-detection in individual bands shortward of $J_{125}$, we use a non-detection criterion following \citet{Bouwens11c} based on $\chi_{opt}^2 = \Sigma_{i} \textrm{SGN}(f_{i}) (f_{i}/\sigma_{i})^2$, with $f_{i}$ the flux in band $i$ and $\sigma_i$ the associated uncertainty. SGN($f_{i}$) is equal to 1 if $f_{i}>0$ and $-1$ if $f_{i}<0$, and the summation runs over all the bands available in a given field shortward of $J_{125}$. Typically these are $B_{435}$, $V_{606}$, $I_{814}$, and $Y_{105}$. We then adopt a criterion $\chi_{opt}^2 < 2.5$. This efficiently excludes lower redshift contaminants while only reducing the selection volume by a small amount \citep[20\%; see also][]{Bouwens15UVLFs,Oesch14}. 

In order to guard our selection against contamination by intermediate redshift, dusty galaxies, we also measure the $H_{160}-[4.5]$ colors of all sources, and exclude galaxies with colors $>2$. No such sources were identified in the HFF field, but a small number of galaxies had been excluded in our previous searches over the GOODS fields based on this criterion \citep[see e.g.,][]{Oesch12a}.

When applying the above selection criteria to all $HST$ fields listed in Table \ref{tab:fields}, we identify nine reliable candidate $z>9.5$ galaxies in total. These sources are discussed in the following sections.

\begin{figure}[tbp]
	\centering
	\includegraphics[width=\linewidth]{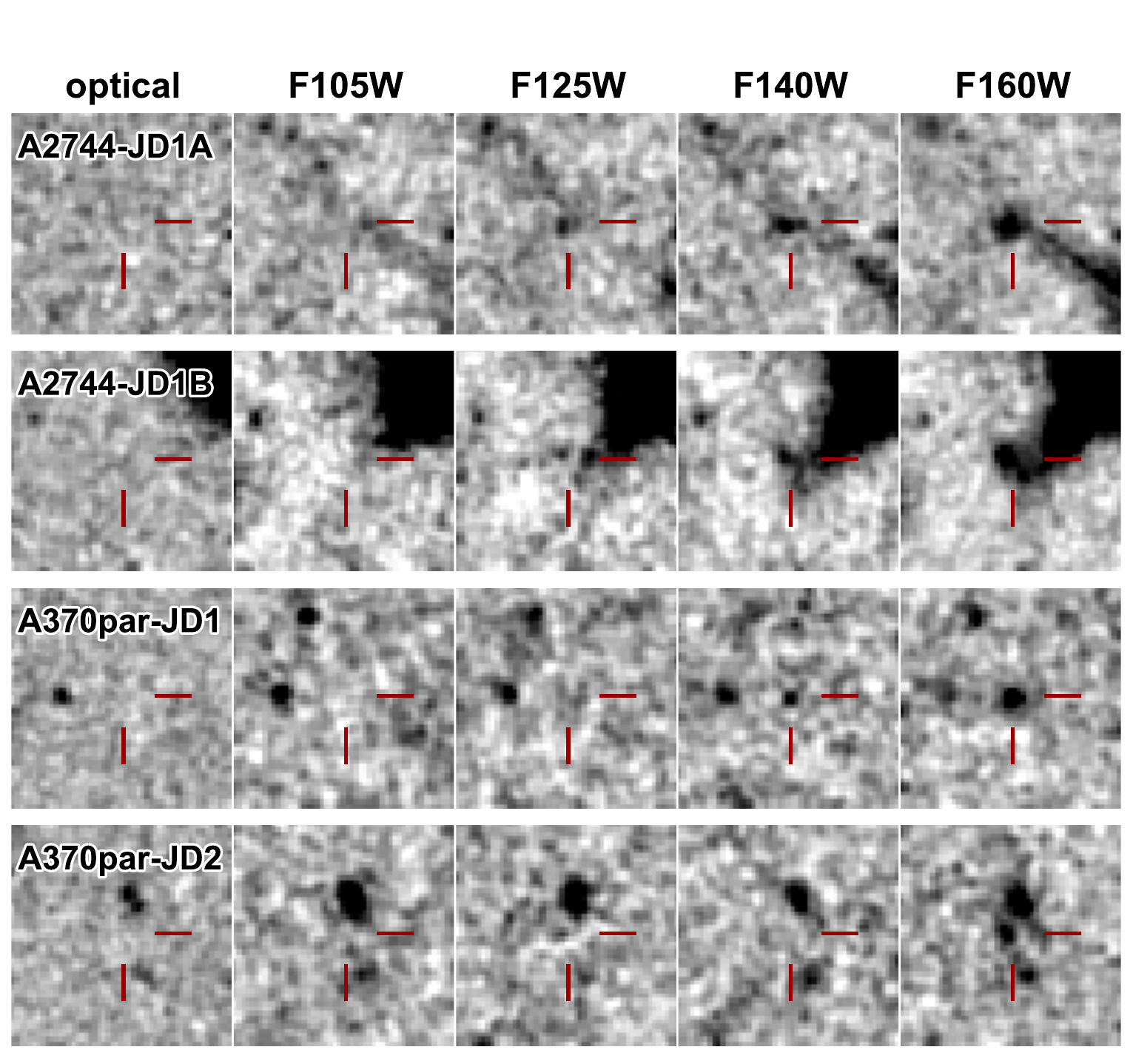}
  \caption{Stamps of the only four $z\sim10$ galaxy candidates with S/N$>6$ in the twelve HFF cluster+parallel fields. From left to right these show a stack of all the available optical ACS images, and the WFC3/IR $Y_{105}$, $J_{125}$, $JH_{140}$, and $H_{160}$ images. The images are oriented North up and they span 3\arcsec on a side. $z\sim10$ candidates are identified based on $J_{125}-H_{160}>1.2$ colors and non-detections at shorter wavelengths.}
	\label{fig:stampszgtr8}
\end{figure}

\subsection{LBG Candidates in the HFF Fields}
\label{sec:HFFcand}

We first discuss the $z\sim10$ candidate sample based on the HFF dataset in detail, since we have so far only published our search result for the first HFF cluster and its parallel field \citep[A2744;][]{Oesch15}. As outlined above, we adopt a strict S/N$>6$ cut for the HFF fields to ensure a reliable candidate selection. Only two of the twelve HFF WFC3/IR fields reveal such candidates in our search. These are the cluster field of Abell 2744 and the parallel field of Abell 370. A few additional, potential $z\sim10$ sources are shown in the appendix together with a note for each, as to why it was not kept in the final candidate list. 

The two candidates behind the cluster Abell 2744 (A2744-JD1A, and A2744-JD1B) were already presented in \citet[][]{Oesch15}. They were originally identified in \citet{Zitrin14} as two images on either side of the critical curve of one intrinsic, very faint source. While the observed magnitude of these sources is $H=26.3$ and 26.8 mag, the expected magnification for these images lies in the range 7 to 27 when considering the full range of all HFF lens models.
However, most models agree on a magnification factor of $\mu=12-13$ for A2744-JD1A. For A2744-JD1B, the distribution of predicted values is bi-modal, clustered around $\mu\sim12$ or $\mu\sim25$. Only the higher values result in a self-consistent intrinsic magnitude in the source plane, however, and they are thus more likely. In particular, the magnification factors from the Glafic v3 lens model that are also shown in Table \ref{tab:phot} reproduce a consistent intrinsic magnitude based on both images of $H_{int}=29.8$ mag. This means that the galaxy is even slightly fainter than the faintest blank-field source found in the HUDF/XDF %(XDFj-38126243) 
\citep[XDFj-38126243][]{Bouwens11a,Oesch13}, which has $H=29.6\pm0.3$.

% with a median magnification of $\mu=12$ for both images. This means the intrinsic magnitude of the lensed galaxy is expected to be $H_{int}\sim29-30$ mag. In particular, the magnification factors from the Glafic v3 lens model reproduce a consistent intrinsic magnitude based on both images of $H=29.8$ mag, i.e., even slightly fainter than the faintest source found in the HUDF/XDF (XDFj38126243), which has $H=29.6\pm0.3$.

As is evident from Fig \ref{fig:stampszgtr8}, the two images lie very close to diffraction spikes of two separate bright foreground sources, which is why photometry for these sources had to be performed manually \citep[see][for details]{Oesch15}. \citet{Zitrin14} further identify a third image of the same source, which is not present in our catalogs, however, as it blended in the halo of a foreground galaxy.
Note that our procedure for computing the UV LF is set up in the image plane, and we thus double-count multiple images, which is why we keep both of these candidates separate.

Interestingly, the two candidates in the parallel field of Abell 370 (A370par-JD1, and A370par-JD2), have a similar observed  $H=27.9$ mag and an H-band S/N of 10.1 and 6.4, respectively. They lie about 1\farcs2 from each other, and are unlikely to be physically associated. While the first candidate is clearly detected in the $JH_{140}$ image, the second source is only seen in $H_{160}$, indicating a significantly higher redshift. The second source also lies close to a pair of foreground sources. The magnification factor for these two sources in the parallel field is provided by the lens model of \citet{MertenModel}, who find a 10$-$16\% residual magnification. The two foreground galaxies in front of A370par-JD2 are not expected to significantly increase this magnification due to galaxy-galaxy lensing given their faintness ($H_{160}=27.3$ and 28.8, respectively) and inferred low mass.

\subsection{Comparison to other HFF $z\sim10$ Galaxy Searches}

Several previous authors have searched for $z\sim10$ galaxy candidates using the HFF data or parts of it. In particular, \citet{McLeod16} used the first four HFF cluster and parallel fields in combination with previous CLASH data to identify $z>8.4$ galaxy candidates based on a photometric redshift selection. In the eight HFF+parallel fields, they present only two candidates in their $z\sim10$ sample, both with $z_{phot}=9.5$. One of these sources (their ID: HFF1C-10-1) corresponds to our candidate A2744-5887. The other source, HFF4P-10-1, is a robust high-redshift candidate. However, given its color $J_{125}-H_{160}=0.86\pm0.26$ it does not satisfy our selection and it likely lies closer to $z\sim9$ than $z\sim10$. Indeed, it was also selected as a $z\sim9$ LBG in \citet[][ID: HFF4P-3994-7367]{Ishigaki17}. 

\citet{Infante15} present one candidate galaxy, ID 8958, in the M0717 cluster field with a photometric redshift of 10.1, which would be the faintest known $z\sim10$ source given its inferred magnification factor of $\mu\simeq20$. This source is also present in our catalogs. However, it only has a S/N of 4.3 in $H_{160}$ and 3.2 in $JH_{140}$, and therefore does not pass our final selection. We confirmed this low S/N by hand through aperture photometry using the iraf task \texttt{qphot}. That said, the candidate has an extended morphology along the shear axis of the magnification and appears likely to be a real source. Nevertheless, deeper data would be required to confirm it as a robust $z\sim10$ candidate, and we do not include it in our analysis. As can be seen later, the extrapolation of our UV LF to the absolute magnitude of this source is consistent with the measurement from \citet{Infante15} within $<1\sigma$, however.

The cluster M1149 contains the $z\sim9.5$ galaxy MACS1149-JD, first reported in \citet{Zheng12} based on CLASH data \citep{Postman12}. The deeper HFF data confirmed the strong Lyman break of the source \citep{Zheng17} and a weak continuum detection in an HST grism spectrum is consistent with the photometric redshift of $z_{phot}=9.5\pm0.1$ \citep{Hoag17}. The source lies just barely outside of our color selection, however, with a measured color of $J_{125}-H_{160}=1.14\pm0.07<1.2$, which is why it is not included in our sample.

Finally, \citet{Ishigaki17} present the results of a high-redshift galaxy search from the full HFF dataset \citep[see also][for an earlier analysis]{Ishigaki15}. Interestingly, even though these authors use effectively identical search criteria to what we use, they do not report any $z\sim10$ galaxy candidates. Since \citet{Ishigaki17} do not subtract the ICL before running their detection algorithm, however, it is not surprising that they do not recover the candidates in the cluster fields. The reason why they do not identify our candidates in the parallel fields are less obvious, but likely involve differences in the PSF homogenization, source deblending, and aperture photometry (versus isophotal photometry for color and S/N measurements).

\subsection{$z\sim10$ Candidates in Other Search Fields}
\label{sec:PrevCand}

The $z\sim10$ searches and galaxy candidates from the remaining $HST$ fields have already been presented in previous papers \citep{Oesch12b,Oesch14,Bouwens16z910}. They are summarized in Table \ref{tab:prevCand}. In particular, the sample consists of five reliable sources identified in the XDF and the two GOODS fields. While the XDF source is found close to the detection limit of the data with $H=29.9$ mag, the four sources identified in the CANDELS/GOODS fields are 3-4 mag brighter with $H=26.0-26.9$ mag. Surprisingly, no candidates were found with intermediate magnitudes, even though the CANDELS-Deep data would have been sensitive down to $H\sim28$ mag (see the discussion in our previous papers).

Additionally, no $z\sim10$ candidates were found in the deep HUDF09 parallel fields, or in the three CANDELS Wide fields. The only exception is a lower-quality candidate in the CANDELS/EGS field which was only partially confirmed by a follow-up HST program \citep[]{Bouwens16z910}. Since it has a 30\% chance to be a lower redshift contaminant based on the derived redshift probability distribution function from an SED fit, we have not included it in the current analysis.

%%%%%

\begin{deluxetable}{lccccc}
\tablecaption{Additional $z\sim10$ Galaxy Candidates from Previous Analyses \label{tab:prevCand}}
\tablewidth{\linewidth}
\tablecolumns{6}
%\tabletypesize{\tiny}

\tablehead{\colhead{ID} & R.A. & Decl. &\colhead{$H_{160}$}  & \colhead{$M_{UV}$} & Ref.}

\startdata

XDFj-38126243  &  03:32:38.12  &  -27:46:24.3  &  $29.6\pm0.3$ & -17.9 & 1,2 \\

GN-z11\tablenotemark{a} & 12:36:25.46  & 62:14:31.4 & $26.0\pm0.1$ &  -21.6 & 3,4,5 \\
GN-z10-2 & 12:37:22.74  & 62:14:22.4 & $26.8\pm0.1$ & -20.7 & 3 \\
GN-z10-3 & 12:36:04.09 & 62:14:29.6 & $26.8\pm0.2$ & -20.7 & 3 \\
GS-z10-1 & 03:32:26.97  & -27:46:28.3 & $26.9\pm0.2$& -20.6 & 3 \\
\cutinhead{Possible Candidate Not Included in Analysis}
EGS910-2\tablenotemark{b}  &  14:20:44.31 & 52:58:54.4 & $26.7\pm0.2$ & -20.8  & 6

\enddata

\tablerefs{(1) \citet{Oesch13}, (2) \citet{Bouwens11a}, (3) \citet{Oesch14}, (4) \citet{Oesch16}, (5) \citet{Bouwens10c}, (6) \citet{Bouwens16z910}  }
\tablenotetext{a}{This source has been spectroscopically confirmed to lie at $z=11.1\pm0.1$ by \citet{Oesch16}. It satisfies our color criteria consistent with the expected redshift distribution function of our LBG selection and is thus included in the following analysis.}
\tablenotetext{b}{This source only has a probability of 71\% to lie at $z>9$, and is thus not yet included, until deeper follow-up observations confirm its high-redshift nature.}

\end{deluxetable}

%%%%%%

\subsection{Galaxy Selection Functions}
\label{sec:completeness}

The computation of a UV LF requires an accurate knowledge of the effective selection volume of a given sample. To derive this, we perform extensive completeness and redshift selection simulations, as we have done in previous analyses. In particular, we insert artificial galaxies into the science data and recover these using the same procedure as has been used for the selection of the actual galaxy candidates.

It is clear from our simulations that the detection completeness depends on the light profile and the size distribution of galaxies. This is particularly important in the very high-magnification regions of the HFF cluster datasets \citep[see e.g.][]{Oesch15,Bouwens16size}. 
\citet{Bouwens17HFF1} show that both the assumptions on galaxy sizes and the uncertainties of the magnification factors can lead to underestimated selection volumes and hence to overpredicted faint-end slopes of the UV LFs based on the HFF cluster dataset.

To simulate a galaxy population that is as close as possible to reality, we adopt simulations that are based on actual light profiles of $z\sim4$ LBGs observed in the XDF field that are cloned to higher redshift and scaled to reproduce the observationally constrained size distribution functions at $z>4$. This technique can only be applied to the blank field datasets, however. In the highest magnification regions of the HFF clusters, the achieved spatial resolution is too high compared to the observed $z\sim4$ galaxy light profiles even when a size scaling with redshift is taken into account. In the case of the HFF clusters, we thus adopt pure Sersic light profiles for the galaxy population \citep[see also][]{Oesch15}.

The adopted size distribution functions are based on recent observations of both the redshift and luminosity dependence of galaxy sizes \citep[see e.g.][]{Oesch10b,Grazian12,Huang13,Ono13,Holwerda15,Kawamata15,CurtisLake16,Bouwens16size}.
In particular, we model sizes with lognormal distributions with $\sigma=0.2$\,kpc, and normalizations that depend both on redshift (following $r_{eff}\propto(1+z)^{-1.5}$) and on luminosity (according to $r_{eff}\propto L_{UV}^{0.5}$). The luminosity dependence is particularly important for the cluster fields where highly magnified galaxies are being found with extremely small sizes \citep{Kawamata15,Bouwens16size,Vanzella17}.

\begin{deluxetable}{cc}
\tablecaption{Stepwise Determination of the $z\sim10$ UV LF Based on all HST Legacy Fields \label{tab:z10lf}}
\tablewidth{0.65\linewidth}
\tablecolumns{2}
\tablehead{$M_{UV}$ [mag] & $\phi_*$  [10$^{-4}$Mpc$^{-3}$mag$^{-1}$]   }

\startdata

$-22.25$  &  $<0.017$\tablenotemark{*} \\ 
$-21.25$  &  $0.010^{+0.022}_{-0.008}$  \\ 
$-20.25$  &  $0.10^{+0.10}_{-0.05}$  \\ 
$-19.25$  &  $0.34^{+0.45}_{-0.22}$  \\ 
$-18.25$  &  $1.9^{+2.5}_{-1.2}$  \\ 
$-17.25$  &  $6.3^{+14.9}_{-5.2}$

\enddata

\tablenotetext{*}{90\% upper limit.}

\end{deluxetable}

Finally, the redshift selection function depends on the UV continuum color distribution of galaxies. We therefore assume a distribution of UV continuum slopes that matches the observed $\beta_{UV}$ values, including its luminosity dependence. 
We assume no redshift evolution in the $\beta_{UV}$ distributions beyond $z>8$ \citep{Wilkins16}, and fix the relation to the one observed at $z=8$ by \citet{Bouwens14b}. These are similar to other estimates \citep[see e.g.][]{Finkelstein12,Dunlop13}.

For each blank field dataset, we insert 100,000 galaxies with the above physical properties and with redshifts ranging from $z=8$ to $z=12$, and we compute the magnitude dependent completeness, $C(m)$, as well as the redshift selection functions, which depend both on redshift and  magnitude $S(z,m)$. The combination of these two quantities allows us to compute the effective selection volume at a given observed magnitude, $m$, as well as the expected number of observed sources for a given model UV LF (see next section). 

As shown, e.g., in \citet{Oesch15}, the completeness also depends on lensing magnification in the HFF cluster fields, with reduced completeness at high magnification factors (since galaxies are not point sources). Following the same procedure of that paper, we thus compute the selection probabilities $p(z,m_{obs}, \mu)$, which depends on the redshift, the observed magnitude $m_{obs}$, and the magnification $\mu$ of the simulated galaxies in the cluster fields.

\begin{figure}[tbp]
	\centering	
	\includegraphics[width=\linewidth]{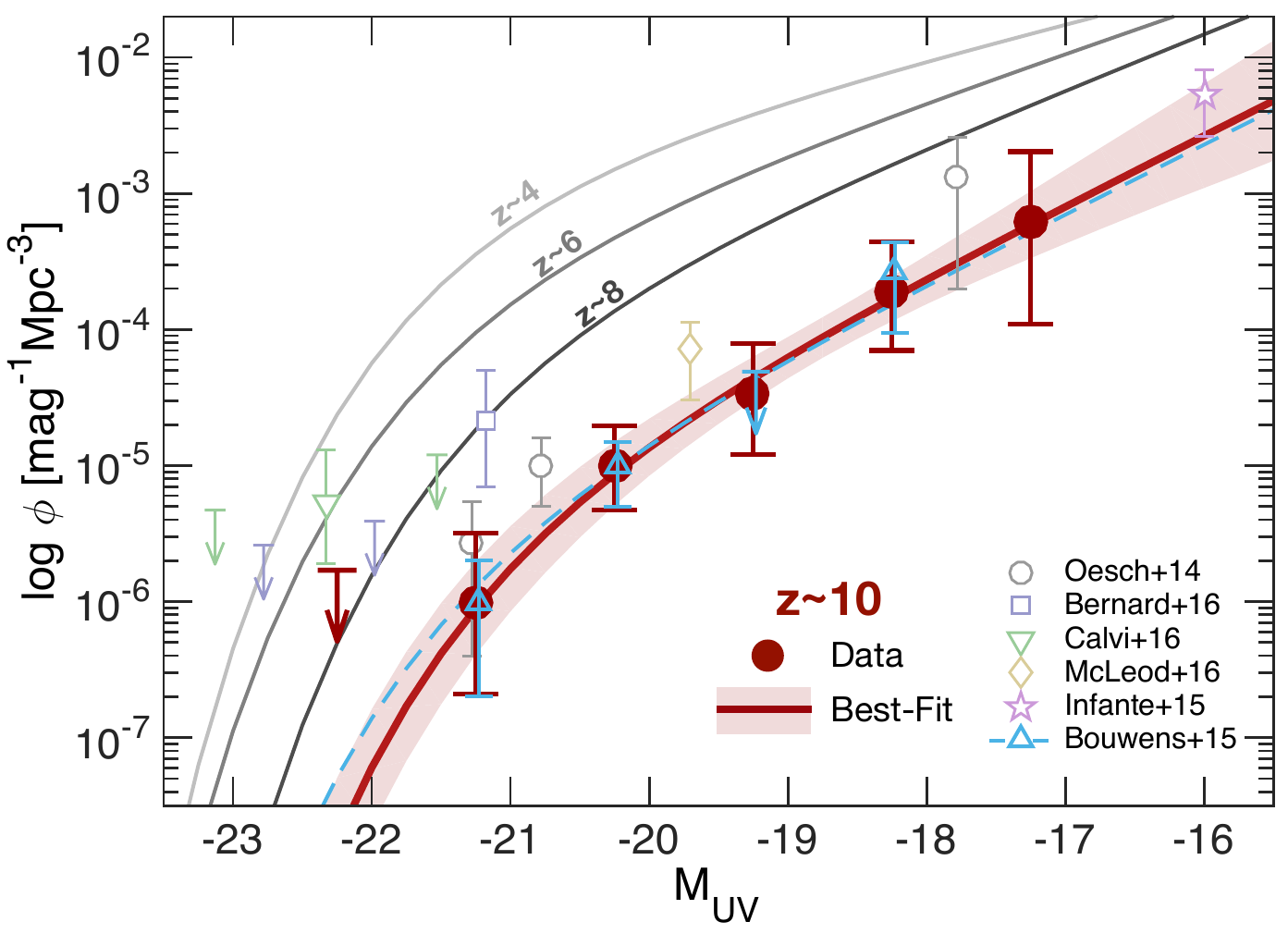}
  \caption{The measured UV LF at $z\sim10$. The red points with errorbars correspond to our new measurements from all the $HST$ legacy datasets. The red solid line is our best-fit LF with $1\sigma$ uncertainty contours. These new results can be compared to previous measurements based on a subset of these datasets from \citet{Oesch14} (gray open circles) and the independent analysis from \citet{Bouwens15UVLFs} (dashed blue line and open triangles). Given the overlap in datasets (5 of the 9 candidates are from the older datasets), the excellent agreement with the latter is not entirely surprising, but it is nevertheless encouraging. The new measurements confirm our previous results and show again that the UV LF evolves very rapidly at $z>8$. Some previous analyses resulted in higher values for the UV LF, however, which include the BORG analyses from \citet{Bernard16} and \citet{Calvi16}, as well as the HFF+CLASH analysis from \citet{McLeod16}. This is discussed more extensively in Section \ref{sec:prevLFs}.
  }
	\label{fig:LFobs}
\end{figure}

\begin{deluxetable*}{lcccc}
\tablecaption{Comparison of $z\sim10$ LF Determinations With Different Assumptions and With Previous Work \label{tab:lfcomparison}}
\tablewidth{0.8\linewidth}
\tablecolumns{5}
\tablehead{Assumption & $\log\phi_*$  [Mpc$^{-3}$mag$^{-1}$]  &  $M_{UV}^*$ [mag]  &  $\alpha$  & $\log$\,SFRD [$M_\odot$/yr]\tablenotemark{$\dagger$} }

\startdata

{\bf Fixed M* (at $z\sim8$ LF)\tablenotemark{*}} & $-4.89^{+0.24}_{-0.30}$   &  $-20.60$ (fixed)  & $-2.28\pm0.32$ & $-3.29\pm0.16$ \\[3pt] 

Density Evolution w.r.t $z\sim8$ & $-4.72\pm0.15$   &  $-20.63$ (fixed)  & $-2.02$ (fixed)  & $-3.31\pm0.15$  \\

Fixed M* (at average from $z\sim4-8$)    & $-5.13^{+0.25}_{-0.28}$   &  $-20.9$ (fixed)  & $-2.33\pm0.30$  & $-3.30\pm0.15$   \\

\cutinhead{\textit{Previous Determinations}}

Bouwens et al.\ (2015) &   $-5.1\pm0.2$     &  $-20.92$ (fixed)   &  $-2.27$ (fixed) & $-3.32^{+0.36}_{-0.45}$\tablenotemark{$\ddagger$}  \\
Oesch et al.\ (2014)  &  $-4.27\pm0.21$     &  $-20.12$ (fixed)   &  $-2.02$ (fixed) & $-3.14\pm0.21$\tablenotemark{$\ddagger$}  \\ 

McLeod et al.\ (2016) &  $-3.90^{+0.13}_{-0.20}$     &  $-20.10$ (fixed)   &  $-2.02$ (fixed) & $-2.78^{+0.13}_{-0.20}$\tablenotemark{$\ddagger$} \\

Ishigaki et al.\ (2017)  &  $-4.60^{+0.14}_{-0.22}$     &  $-20.35$ (fixed)   &  $-1.96$ (fixed) & $-3.38^{+0.14}_{-0.22}$\tablenotemark{$\ddagger$}  \Bstrut

\enddata

\tablenotetext{*}{This is used as our best-fit model throughout the rest of the paper.}
\tablenotetext{$\dagger$}{Integrated down to $M_{UV}=-17.0$, corresponding to a SFR of 0.3 $M_\odot$\,yr$^{-1}$, based on the conversion factor from UV luminosity to SFR: ${\cal{K}}_{UV}=1.15\times10^{-28}$ $M_\odot$\,yr$^{-1}/$erg\,s$^{-1}$\,Hz$^{-1}$}
\tablenotetext{$\ddagger$}{Computed based on the Schechter function parameters, using a consistent conversion factor ${\cal{K}}_{UV}$ and integration limit $M_{UV}=-17.0$ to allow for a quantitative comparison.}

\end{deluxetable*}

\section{Results}
\label{sec:results}

\subsection{The UV LF at $z\sim10$}

In the following sections we first present our observational constraints on the UV LF at $z\sim10$ based on the combination of all legacy HST fields, and we then compare these to model predictions. 

\subsubsection{Observational Constraints}

Using the $z\sim10$ candidate sample described in sections \ref{sec:HFFcand} and \ref{sec:PrevCand}, as well as the completeness and selection functions discussed in section \ref{sec:completeness}, we can now estimate the UV LF at $z\sim10$. We compute step-wise UV LFs by binning the galaxy population in absolute magnitudes $M_{UV}$ and estimating the effective volume at that magnitude. 
For each blank field dataset $i$, this can be written as:
\begin{equation}
V_{eff,i} = \int dz ~ S_i(z,m[M_{UV},z])C_i(m[M_{UV},z]) \frac{dV}{dzd\Omega} d\Omega_i
\end{equation}
Where $d\Omega_i$ corresponds to the survey area of field $i$. The total effective volume at a given absolute magnitude is then simply the sum over all survey fields. A similar equation holds for the cluster fields, except that the volume element $dV$ and the relation between the observed magnitude and the intrinsic absolute magnitude depend on the magnification factor $\mu$ \citep[see, e.g.,][]{Oesch15}.

Uncertainties on the step-wise LF points are derived using Poisson statistics, which dominate the error budget at low number counts even though cosmic variance is significant for the small individual survey fields we probe here. We include the effect of cosmic variance by adding a fractional uncertainty on the expected number counts per field which ranges from 35\% in the CANDELS-Wide fields up to 65\% in the HFF cluster fields \citep[e.g.][]{Trenti08,Robertson14}.

The resulting step-wise UV LF is shown in Fig. \ref{fig:LFobs} and is tabulated in Table \ref{tab:z10lf}. As the figure shows, the data points lie significantly below the $z\sim8$ LF, confirming earlier claims for a significant evolution across $z=8-10$. The LF is in excellent agreement with our previous analysis from  \citet{Bouwens15UVLFs}, but the HFF candidates nicely fill in the one upper limit that was still present in earlier LF determinations at $M_{UV}\sim-19$. The combined dataset now contains candidates spanning a full 4 mag range.

In order to describe the UV LF, we adopt a Schechter function \citep{Schechter76}, whose parameters are fit to the observed number of galaxies in bins of observed $H_{160}$ magnitudes.
Note that the resulting Schechter function is thus not derived from a fit to the stepwise LF, but from an independent fit to the detected number of sources in observed magnitude bins. This is much more accurate as it allows us to properly take into account the redshift distribution as a function of magnitude and the K-correction as a function of redshift. In particular, for a given model UV LF $\phi$, we compute the expected number of galaxies:
\begin{equation}
dN_{exp}/dm = \int dz ~ S(z,m)C(m) \phi(M_{UV}(m,z)) \frac{dV}{dzd\Omega} 
\label{eq:Nexp}
\end{equation}

To derive the Schechter function parameters, we then maximize $\cal{L} $ $= \prod_{i} \prod_j P(N^{\rm obs}_{i,j},N^{\rm exp}_{i,j})$, where $P$ is the Poissonian probability,  $i$ runs over all fields, and $j$ runs over the different magnitude bins. Due to the small number of galaxy candidates and the strong degeneracies between the Schechter function parameters, an unconstrained, simultaneous fit of all three parameters is not yet meaningful. We therefore perform different fits with various parameters kept fixed at values motivated by $z\sim6-8$ UV LFs.
The results are listed in Table \ref{tab:lfcomparison}. 

Of particular importance are fits where the characteristic UV luminosity is kept fixed at lower redshift values given the recent finding of very little evolution in this parameter across $z\sim4$ to $z\sim8$. We test two possibilities, one where we fix $M_*$ at the average value found at $z\sim4-8$ from both \citet{Bouwens15UVLFs} and \citet{Finkelstein15}, i.e. $M_*=-20.9$. Using this constraint we find a slightly lower normalization $\phi_*$ and a somewhat steeper faint end slope $\alpha$ than fixing $M_*=-20.6$, which is the value of the $z\sim8$ LF found in \citet{Bouwens15UVLFs}. However, these two LFs are effectively indistinguishable from each other over the luminosity range we probe here. Indeed, their inferred SFRDs are in excellent agreement, as can also be seen in Table \ref{tab:lfcomparison}. In fact, all three of our LFs give essentially identical SFRDs.

We also list the Schechter function based on a pure density evolution from $z\sim8$ to $z\sim10$ relative to the $z\sim8$ UV LF parameters from \citet{Bouwens15UVLFs} (see second row in Table \ref{tab:lfcomparison}), which results in almost exactly an order of magnitude (1.04 dex) lower normalization at $z\sim10$ than at $z\sim8$.
Throughout the rest of the paper, we will use the UV LF with $M_*=-20.6$, $\log \phi_* = -4.89^{+0.24}_{-0.30}$ Mpc$^{-3}$mag$^{-1}$ and $\alpha=-2.28\pm0.32$ as our best fit model.

\begin{figure*}[tbp]
	\centering	
	\includegraphics[width=0.89\linewidth]{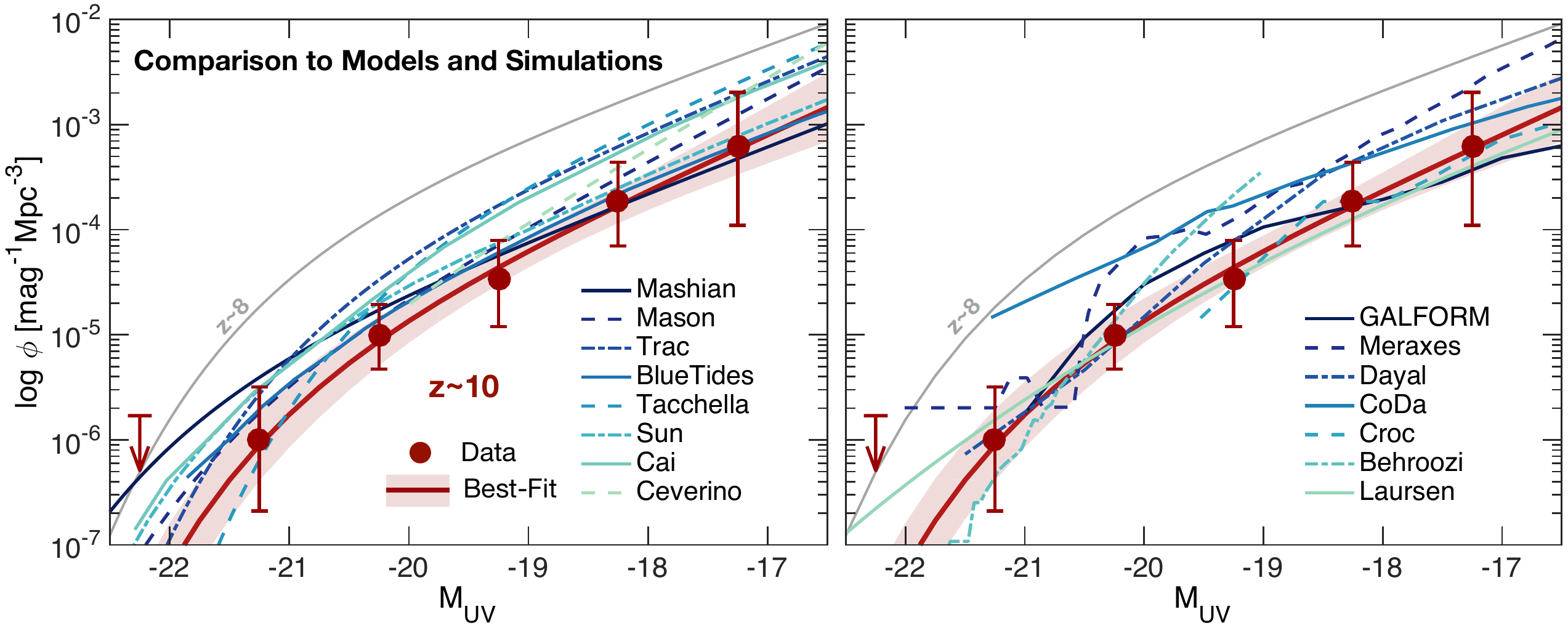}
  \caption{The $z\sim10$ UV LF results compared to models and simulations from the literature. The red line and points show our observed LF, while the blue to green lines are model LFs (see text for references). All models predict that the UV LF evolves significantly from $z\sim8$ (light gray line) to $z\sim10$. They typically do not differ by more than a factor of 3-4 over the luminosity range of the data. While several simulated UV LFs are in good agreement with the observations and generally within the $1\sigma$ range, our observed UV LF lies at the lower end of essentially all theoretically predicted LFs.
  }
	\label{fig:LFcompare}
\end{figure*}

\subsubsection{Comparison to Previous Measurements}
\label{sec:prevLFs}

Several early estimates of the UV LF at $z\sim10$ have been published in the past. However, these have all been based on much smaller search areas than studied here or were based on a subset of the data analyzed in this paper. In Figure \ref{fig:LFobs}, we show several of these previous estimates. In particular, these include our own measurements based on the analysis of only the two CANDELS/GOODS+HUDF fields \citep{Oesch14}. While these earlier measurements are consistent within the uncertainties, the fully combined $HST$ dataset now indicates an even lower normalization by a factor $\sim2\times$ compared to the lowest of these earlier estimates. In a large part, this is due to the fact that the GOODS-North field appears to contain a significant overdensity of luminous ($M_{UV}<-20$ mag) galaxies, with three galaxies within a small projected area. Even though the remaining CANDELS-Wide fields would have reached faint enough to find such sources, no reliable candidate could be identified in these datasets, resulting in the lower inferred number density. 

Our measured UV LF is in excellent agreement with the previous analysis by \citet{Bouwens15UVLFs}. This is very encouraging. Our analyses are completely independent, including the simulations of the selection functions and the candidate searches. However, this result is also not surprising given the fact that we use a very similar approach and the fact that the datasets largely overlap, with the exception of the HFF dataset which we newly analyze here. As can be seen, the HFF candidates now provide a measurement of the UV LF at $M_{UV}=-19.25$. This magnitude bin previously did not contain any galaxies and corresponded to an upper limit. 

Even though the Schechter function parameters of our best-fit model are very different from the ones quoted in \citet[][see also Table \ref{tab:lfcomparison}]{Bouwens15UVLFs}, the UV LFs are in very good agreement with each other over the luminosity range we probe, as is shown in Figure \ref{fig:LFobs}. Similarly good agreement is found with the Schechter function from \citet{Ishigaki17}. 

The only previous UV LF that is clearly discrepant from our new measurement is the one from \citet{McLeod16}. This is based on \textit{only one} UV LF point, shown in Fig \ref{fig:LFobs}, which is $\sim3\times$ higher than our measurement. Based on this one point, \citet{McLeod16} argued for a much more gradual decline in the cosmic SFRD at $z>8$. While we discuss the evolution of the cosmic SFRD in detail in section \ref{sec:SFRD}, it is worth mentioning here that the combination of all the $HST$ legacy fields is clearly inconsistent with that UV LF (and hence the SFRD) from \citet{McLeod16}. In particular, we can compute how many $z\sim10$ galaxies we would have expected in our combined dataset assuming their UV LF (based on equation \ref{eq:Nexp}). This calculation results in 28 galaxies, i.e., a factor $\sim3\times$ larger than our actual sample of only 9 sources, and can robustly be ruled out. In particular, the \citet{McLeod16} LF predicts $\sim2$ galaxy candidates per HFF cluster field and 4.6 galaxies in the six HFF parallel fields, meaning that in the HFF dataset alone we should have found 16.7 $z\sim10$ galaxy candidates, four times as many as are present in the data.

Using our best-fit LF, we find much better agreement between the observed and predicted numbers. In particular, this LF predicts a total of 3.3 and 1.3 galaxies to be found in the six HFF clusters and parallel fields combined, respectively -- in good agreement with the four candidate images we actually identified.

\subsubsection{Comparison to Predicted UV LFs from Models}
\label{sec:ModelLFcomparison}

Given the small number of $z\sim10$ galaxies that we identified in the combined HST dataset, it is interesting to ask whether this might be an indication of a reduced star-formation efficiency in early halos at $z>8$. To this end, we compare our observational results with several theoretical models and simulations of the UV LF evolution that have been published in the literature over the last few years. These include semi-empirical models tuned to the lower redshift LBG LFs or MFs \citep[e.g.,][]{Tacchella13,Cai14,Mason15,Trac15,Behroozi15,Sun16,Mashian16}, or semi-analytical models \citep[e.g.][]{Dayal14,Cowley17}, and full hydrodynamical simulations. The latter include UV LFs from the \textit{CoDa} simulation \citep{Ocvirk16}, the \textit{Croc} simulation suite \citep{Gnedin16}, \textit{BlueTides} \citep[][]{Liu16,Wilkins17}, and \textit{DRAGONS/Meraxes} \citep{Liu16}, the First Light Project \citep{Ceverino17}, as well as another simulation by Laursen et al. (2018, in prep).

The comparison to these theoretical predictions is shown in Figure \ref{fig:LFcompare}. It is evident that the modeled UV LFs decrease in a similar way at $z>8$ compared to what we are finding observationally. Given the vastly different nature of these models, it seems clear that the main driver for this strong evolution is the underlying dark matter halo mass function, which is known to evolve rapidly at these early times (see also discussion in Section \ref{sec:DMHaloBuildup}). The differences among the model predictions is typically less than a factor 3-4 over the luminosity range probed by our observations. 

Overall the agreement between the observed and simulated UV LFs at $z\sim10$ is excellent. Several models lie within the $1\sigma$ uncertainty range of our observed best-fit LF over most of the luminosity range of interest. \textit{However, it is important to note that our inferred best-fit Schechter function lies at the low end of all predicted LFs.} We further investigate the significance of this by computing the expected number of $z\sim10$ galaxy candidates that would have been found in the data for each model LF. This is done by folding the model UV LFs through the actual completeness and selection functions of all the different fields (using equation \ref{eq:Nexp}) and summing the numbers.

\begin{figure}[tbp]
	\centering	
	\includegraphics[width=0.99\linewidth]{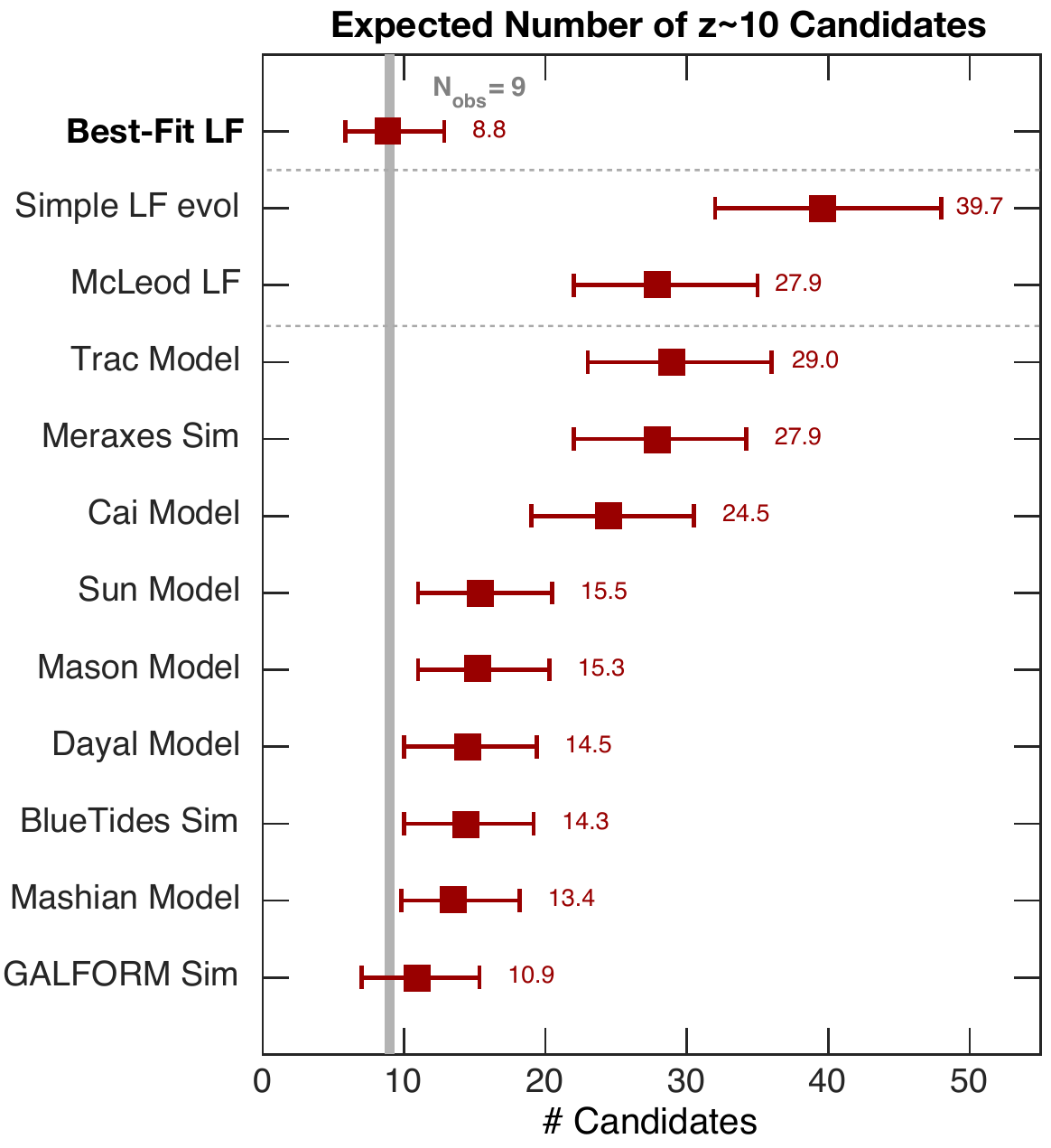}
  \caption{The expected number of $z\sim10$ galaxy candidates in our total imaging dataset based on different models for the UV LF at $z\sim10$ and using the actual selection functions of our search fields. The gray vertical line corresponds to the 9 candidates that we identified. The dark red squares with (Poissonian + Cosmic Variance) errorbars correspond to the predicted number of sources in our survey for different model LFs (only LF models spanning the full luminosity range probed by our survey are shown).
  We can rule out a few of these models that predict more than 25 galaxies. For several other models, our observed number counts are only about 1-1.5$\sigma$ below the predicted numbers that range between 11-16, though they are all consistently high. The model that best reproduces our observed counts comes from the GALFORM simulation \citep{Cowley17}.
  Our $z\sim10$ sample was obtained from the most comprehensive study to date and the low numbers revealed here confirm that a rapid change is occurring between $z\sim8$ and $z\sim10$, faster than across $z\sim4-8$ (accelerated evolution at $z>8$). This can be appreciated from the `Simple LF evol' point, which is just an extrapolation of the LF evolutionary trends across $z\sim4-8$ to $z\sim10$.
   The model trends are broadly in agreement, as seen Fig \ref{fig:LFcompare} and here, but there is still some tension with most models given the small numbers and generally lower LFs (see Section \ref{sec:ModelLFcomparison} for more discussion).
  }
	\label{fig:NExpSim}
\end{figure}

The results of this calculation can be compared to the real number of detected $z\sim10$ candidates, i.e. 9, which is shown in Figure \ref{fig:NExpSim}. Interestingly, there seem to be two classes of models. A small number of models predict that about 25-29 $z\sim10$ galaxies should have been detected. By comparison to the best-fit LF in Fig. \ref{fig:LFcompare}, it is clear that this discrepancy arises mainly due to a higher normalization of the LF at the faint end ($M_{UV}\gtrsim-19.5$). These models appear to form stars in lower mass galaxies too efficiently and thus overpredict the number of galaxies we should have found in the deepest datasets, i.e., the HUDF/XDF and HFF clusters. 

The other set of models predicts galaxy number counts of 13-16, i.e., about 45-70\% higher than the observed 9 candidates. Considering the large Poisson+CV uncertainties on these numbers, these models are all within 1-1.5$\sigma$, and the discrepancies are not significant in each individual case. However, the consistently larger predicted numbers suggest a small disconnect with the observations.

The only model that is in close agreement with the observed number counts and also with the observed LF is the one based on GALFORM presented in \citet{Cowley17}. This model predicts about two magnitudes of extinction in the brightest sources, however, which would result in very red predicted UV continuum slopes ($\beta_{UV}\sim-1.4$ to $-1.1$). In marked contrast to this, the brightest currently known  $z\sim8-10$ galaxy candidates show significantly bluer slopes of $\beta = -2.1\pm0.3$ based on the combination of HST+Spitzer photometry \citep[][see also Oesch et al.\ 2014, Bouwens et al.\ 2014]{Wilkins16}. It will thus be important to test such models with other measurable quantities.

In summary, the comparison of our observations with theoretical predictions shows that our current $z\sim10$ candidate galaxy sample lies at the lower end of the predicted range, both for the model UV LFs and in terms of the absolute number of $z\sim10$ candidates that are present in the data. It will be important to keep this in mind when using these models to define survey strategies for $JWST$ and when predicting higher-redshift number counts.

\begin{figure*}[tbp]
	\centering	
 \includegraphics[width=0.735\linewidth]{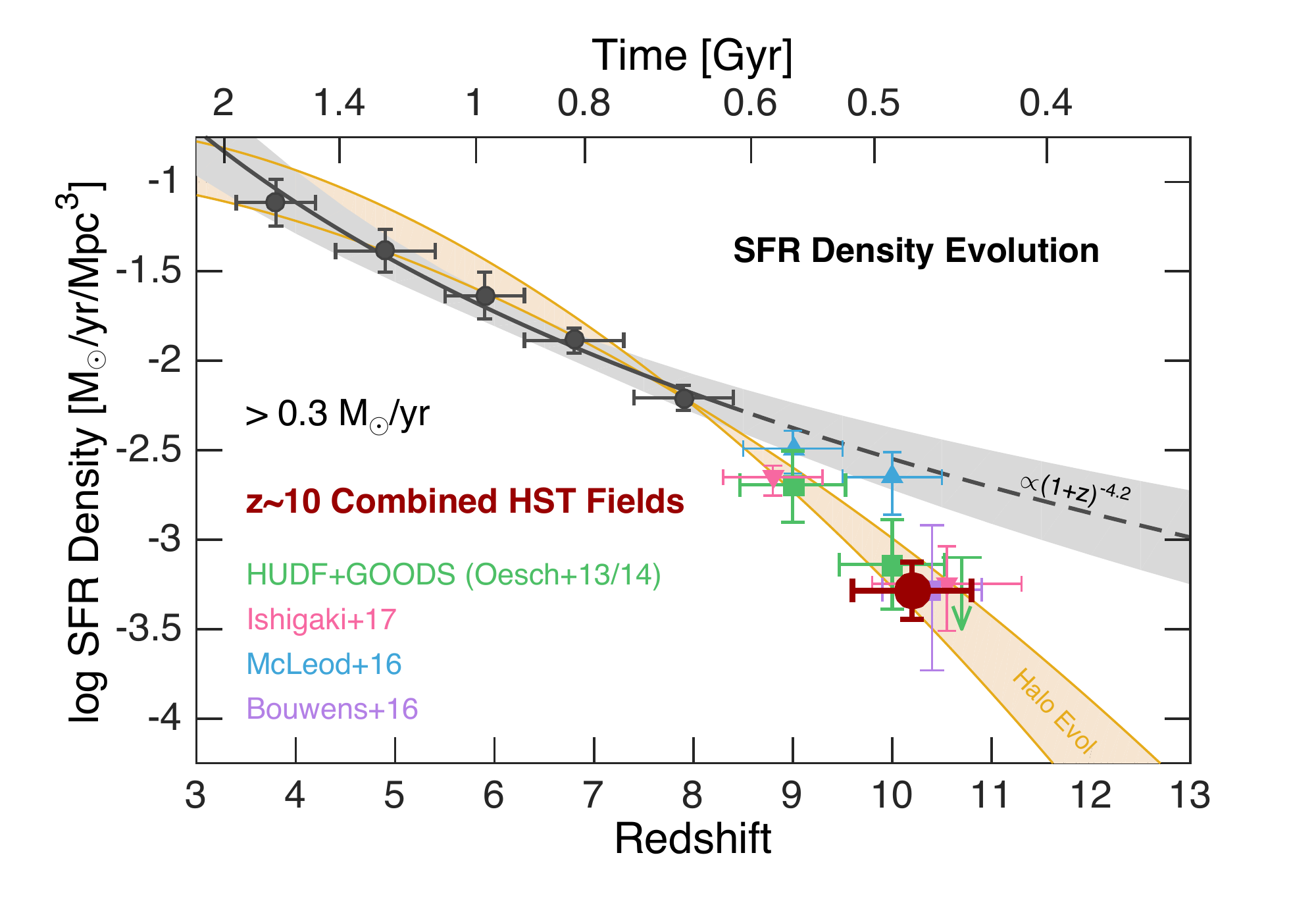} 
% \vspace*{-1.0 cm}
 \caption{The evolution of the cosmic SFR density at high redshift integrated down to a UV luminosity of $M_{UV}=-17.0$ (SFR $>0.3~M_\odot$\,yr$^{-1}$). Lower redshift measurements (gray dots) have been updated from ALMA constraints in \citet{Bouwens16ALMA} assuming an evolving dust temperature and they include a small contribution from ULIRGS. Also shown are the measurements from \citet[][pink triangle]{Ishigaki15} and \citet[][blue triangles]{McLeod16}, which have been corrected to our integration limit and UV luminosity to SFR conversion factor.
 The new measurement from the combination of all $HST$ fields (filled dark red circle) confirms the rapid, accelerated evolution of the SFRD between $z\sim8$ to $z\sim10$, as has previously been found in the HUDF+GOODS fields \citep{Oesch14,Bouwens15UVLFs}. The evolution is significantly faster than at lower redshift (gray shaded region), which is not unexpected given the fast evolution of the halo mass function over this redshift range (see Sect \ref{sec:DMHaloBuildup}). To illustrate this, the orange shaded region shows the relative evolution of the cumulative DM halo mass function integrated down to $\log\, M_h/M_\odot = 9.5-10.5$ and normalized at the $z\sim8$ SFRD value. Clearly, the evolution of the DM halos is in very good agreement with the substantial decrease in the SFRD at $z>8$.
 }
   \label{fig:SFRD}
\end{figure*}

\subsection{Clustering of $z\sim10$ Galaxies}
\label{sec:Clustering}
One possible reason for the low abundance of $z\sim10$ galaxies in the current sample is obviously cosmic variance. 
An indication for a very high bias and clustering strength of bright $z\sim10$ galaxies is provided by the fact that all our nine candidates have been found in only four regions of the sky, while we have searched over 10 general areas. While the CANDELS/GOODS fields have the best data of all the CANDELS fields, the candidates with $H<27$ mag, should have been detectable in any CANDELS field. Yet, only one {\it possible} such source was found in the EGS \citep[see][]{Bouwens16z910}. Similarly, only two of the HFF fields contained any candidates, of which one is a multiple imaged source, while the other field (A370-par) contained two sources. 

This points to the fact that current surveys may simply be too small or not deep enough to provide an accurate sampling of the $z\sim10$ galaxy population. It will thus be crucial to obtain deep, wide-area NIR data over the next few years. This will not be possible with {\it JWST}, which is not a survey telescope. It will either require a significant investment of {\it HST} time or it will have to await new space telescopes such as {\it Euclid} or {\it WFIRST}.

\subsection{The Cosmic SFRD at $z\sim10$}
\label{sec:SFRD}

The evolution of the cosmic SFRD at $z>8$ has been a matter of debate in the recent literature. In particular, several authors claimed a shallower evolution than has been inferred from the combination of the XDF and GOODS datasets by our team. However, most of these studies were based on the analysis of individual, small fields, and an even smaller number of candidates than studied here. Given the large survey volume in the combined HST dataset, we can now establish the best possible constraint on the SFRD at $z\sim10$ based on the UV LFs we derived in the previous section.

Thanks to the lensing magnification in the HFF cluster fields, we have further constrained the UV LF to fainter limits than possible with the HUDF/XDF dataset, allowing us to derive the SFRD to lower limits than in our previous analyses without any extrapolation.  We use an updated conversion factor from UV luminosity to star-formation rate as discussed in \citet{Madau14}: ${\cal{K}}_{UV}=1.15\times10^{-28}$ $M_\odot$\,yr$^{-1}/$erg\,s$^{-1}$\,Hz$^{-1}$. We then integrate the UV LF down to $M_\mathrm{UV}=-17$, which corresponds to a SFR limit of $0.3\, M_\odot$\,yr$^{-1}$, given this adopted conversion factor ${\cal{K}}_{UV}$.

The resulting SFRD values at $z\sim10$ based on the different assumptions about the UV LF Schechter function parameters are tabulated in Table \ref{tab:lfcomparison}. In particular, our best-fit UV LF results in a SFRD value of $\log\dot{\rho_*}=-3.29\pm0.16$ M$_\odot$\,yr$^{-1}$\,Mpc$^{-3}$. This is in very good agreement with our previous measurements, where we already pointed out the accelerated evolution at $z>8$ \citep{Oesch14,Bouwens16z910}. As can be seen from the table, the SFRD also does not change significantly between our different assumptions about the Schechter function parameters. We consistently find values around $\log\dot{\rho_*}=-3.3$ M$_\odot$\,yr$^{-1}$\,Mpc$^{-3}$.

It is interesting to compare this measurement to the SFRD at lower redshift. Fig. \ref{fig:SFRD} also shows these measurements based on new dust correction factors motivated by ALMA observations and adding a small contribution from dusty galaxies. In particular, we plot the values assuming an evolving dust temperature from Table 10 in \citet[][]{Bouwens16ALMA}, which were integrated to the same UV luminosity limit. All numbers were adjusted slightly to account for the different conversion factor ${\cal{K}}_{UV}$. 

This comparison shows that the SFRD (when integrated to a limit of $0.3\, M_\odot$\,yr$^{-1}$) increases by 1.1$\pm$0.2 dex from $z\sim10$ to $z\sim8$. A power law fit to the $z\sim4-8$ values results in a SFRD evolution $\propto(1+z)^{-4.2}$. When extrapolating this to $z\sim10$, our measurement lies a factor 5-6$\times$ below this trend, similar to our earlier findings, but in contrast to some recent claims by other authors \citep[e.g.,][]{McLeod16}. As noted earlier, the previous measurements of the $z\sim10$ SFRD that found values consistent with a simple extrapolation of the lower redshift evolution were all based on very small samples or on very limited search volumes. For example, the SFRD measurement by \citet{McLeod16} was only based on one single point in the UV LF (also shown in Fig \ref{fig:LFobs}), and did not include any constraints from the wider area CANDELS data. The combination of all the $HST$ legacy fields in our analysis significantly increases the search volume and the robustness of the SFRD measurement.

It is also interesting to ask whether a less dramatic evolution of the SFRD is found when integrating to fainter galaxies below our current detection limits. This is indeed the case, given the steeper faint end slope of our best-fit UV LF compared to the $z\sim8$ reference from \citet{Bouwens15UVLFs}. However, even when integrating to $M_{UV}=-13$, i.e. 4 mag below our current detection limit, we still infer a $5-6\times$ lower SFRD at $z\sim10$ compared to $z\sim8$, albeit with large uncertainties due to the significant extrapolation.

\subsection{The $z>8$ SFRD Trend and Dark Matter Halo Buildup}
\label{sec:DMHaloBuildup}

When compared to the SFRD at $z\sim8$ our measured $z\sim10$ value lies almost exactly an order of magnitude lower, given that the Schechter function normalization is found to drop by this amount. Such a fast evolution of the SFRD in only 170 Myr from $z\sim8$ to $z\sim10$ may sound extreme. However, it is important to note that the halo mass function also evolves extremely fast over this same redshift range.
To illustrate this, we compute the cumulative number density of dark matter (DM) halos as a function of redshift using the code \texttt{HMFcalc}\footnote{\url{https://github.com/steven-murray/hmf}} \citep{Murray13}. Our integration limit of the SFRD (0.3 $M_\odot$/yr) roughly corresponds to a dark matter halo mass of $\sim10^{10}$\,$M_\odot$ at $z\sim6-10$ according to several models \citep[e.g.,][]{Mason15,Mashian16,Sun16}. To allow for some variation, we thus compute the cumulative number densities of halos down to $\log\,M_h/M_\odot = 9.5$ and 10.5 as a function of redshift. We then normalize these densities to the measured SFRD at $z\sim8$ to compare the evolutionary trends, which is also shown in Fig. \ref{fig:SFRD}. 

The evolution of the cumulative halo mass functions is in excellent agreement with the fast build-up of the SFRD between $z\sim10$ and $z\sim8$, as is evident from the figure. This shows that the observed fast evolution of the SFRD should not have come as a surprise. It is consistent with a model in which the star-formation efficiency is not varying at high redshift \citep[see also][]{Tacchella13,Mashian16,Stefanon17a}.
Or said another way, a shallower evolutionary trend would require a significant change in the star-formation efficiency within DM halos at high redshift -- to zeroth order.
Recent results from clustering measurements also provide no compelling evidence of such a change \citep[see][]{Harikane17}. 

\begin{figure}[tbp]
	\centering	
 \includegraphics[width=\linewidth]{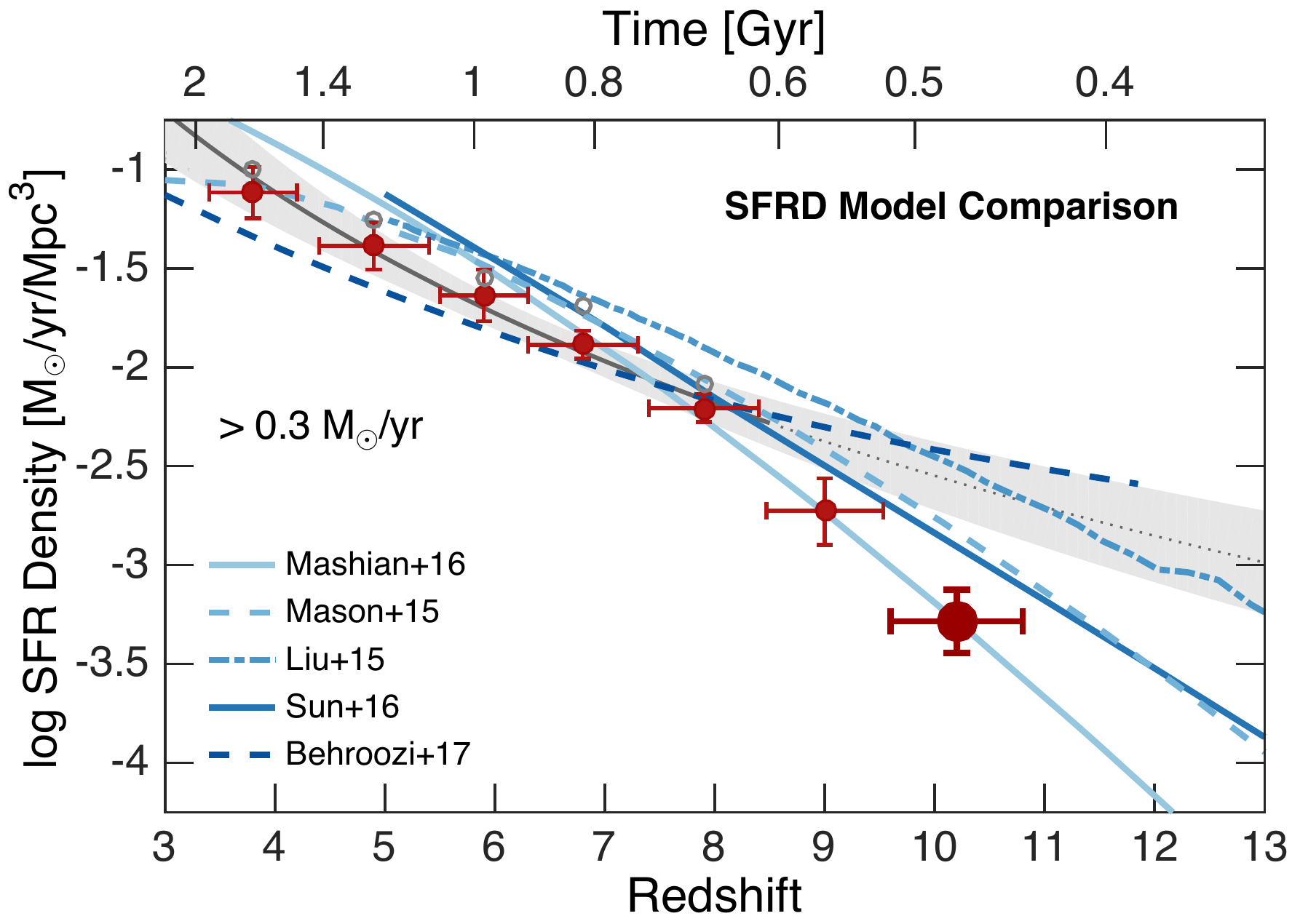} 
% \vspace*{-1.0 cm}
 \caption{The measured SFRD evolution compared model predictions. The red dots are the same measured values as shown in the previous figure corresponding to the results from \citet{Bouwens16ALMA} at $z\sim4-8$, \citet{Oesch14} at $z\sim9$ and this work at $z\sim10$. The model SFRDs (integrated to the same limiting SFR) are shown in blue. They correspond to a subset of the same models shown in the UV LF plot. Similar to the UV LF result, we find that models generally overpredict the SFRD at $z\sim10$ compared to the observations. However, most models do find the SFRD to evolve very rapidly at these epochs and the higher values at $z\sim10$ could partially be explained by a different normalization at $z\sim4-8$. The open gray circles at lower redshift indicate previous measurements that used different dust correction factors \citep[see][]{Bouwens16ALMA}, which were used by most models to tune their parameters. Nevertheless, the predicted SFRDs start to diverge significantly at $z>8$ such that it should be very easy to test and rule out some of these models with early JWST observations.
 }
   \label{fig:SFRDmodels}
\end{figure}

Clearly, our calculation above is simplistic, and more detailed modeling would be valuable, though that is out of the scope of this paper. 
In Figure \ref{fig:SFRDmodels}, we therefore compare the observed SFRD to predictions by some of the same models that were used for the UV LF comparison in section \ref{sec:ModelLFcomparison}, and for which the SFRDs at $z>8$ have been published. In agreement with the UV LF results, most models predict a significantly higher SFRD at $z\sim10$ than we observe. Consistent with our findings that the SFRD evolution can be reproduced simply by the DM halo MF build-up, the model by \citet{Mashian16}, which is based on a fixed SFR-M$_{halo}$ relation at all redshifts reproduces the observed SFRD trends at $z>8$. 
Other models predict higher SFRDs at $z>8$, and the discrepancy between these predictions are increasing with redshift. As is evident from the Figure, by $z\sim12$ the predicted SFRD range spans a factor 30 already. This clearly highlights the great power that the JWST will have in testing these model predictions in the near future.

\section{Summary}
\label{sec:summary}

This paper presented a complete and self-consistent analysis of all the prime extragalactic $HST$ datasets to constrain the evolution of the galaxy UV LF and cosmic star-formation rate densities to $z\sim10$. In particular, the imaging data analyzed here span more than 800 arcmin$^2$, and include the HUDF/XDF, the HUDF09/12, the HFF data, as well as all the five CANDELS fields. In particular, we present a small sample of new $z\sim10$ galaxy candidates in the HFF dataset (see Section \ref{sec:HFFcand}), which are self-consistently folded into the analysis while taking special care of the position- and magnification-dependent completeness in the cluster fields due to lensing. 

The main goal of this paper was to exploit all the available $HST$ Legacy fields in order to derive the best-possible UV LF measurement at $z\sim10$ before the advent of $JWST$.
While it is clear that the UV LF and the cosmic SFRD continue to decline with increasing redshift, there was some debate in the literature on exactly how fast this decline is between $z\sim8$ and $z\sim10$ -- accelerated or continuous with respect to $z\sim4-8$? The HFF dataset provide an excellent, independent test of this, as a power law extrapolation of the UV LF trends to $z>8$ was predicted to reveal more than 25 galaxies in the six HFF clusters+parallel fields. However, after a careful search for $z\sim10$ galaxies based on the Lyman break, our analysis only revealed four reliable candidate sources/images in these HFF fields, clearly demonstrating that the evolution of the UV LF is faster than at lower redshift, confirming our previous findings of an accelerated evolution.

When combining all the HFF candidates with the five sources previously identified in the HUDF+CANDELS data, we indeed find a best-fit UV LF that is almost exactly a factor $10\times$ lower in normalization compared to $z\sim8$ (see Fig. \ref{fig:LFobs} and Table \ref{tab:lfcomparison}). This is consistent with several models of galaxy evolution at the $1-1.5\sigma$ level, as we show in Fig. \ref{fig:LFcompare}. However, the measured UV LFs lie at the low end of all the predictions. In particular, when running the simulated LFs through our selection functions, all models predict a higher number of $z\sim10$ sources than we find observationally (Fig \ref{fig:NExpSim}).

Given this measurement of the UV LF, we compute the cosmic SFRD at $z\sim10$ (integrated over SFRs greater than 0.3 $M_\odot$/yr) and confirm our previous measurement of $\log\dot{\rho_*}=-3.3$ M$_\odot$\,yr$^{-1}$\,Mpc$^{-3}$. This also confirms that the SFRD decreases extremely rapidly at $z>8$, much faster than the evolution seen at $z=4$ to $z=8$ \citep[hence the term ``accelerated evolution"; see][]{Oesch12a,Oesch14}. The rapid decline is consistent with the build-up of the DM halo mass function (see Fig \ref{fig:SFRD}). 

Despite using all $HST$ legacy fields, only a very small number of reliable $z\sim10$ galaxies could be identified. This clearly shows that we are reaching the limit of what is possible with $HST$ to reveal the first generations of galaxies. The discussion about the UV LF at $z\sim10$ is definitely far from settled - in particular, it is striking that the highest redshift object in our sample (GN-z11) is also the most luminous, which hints at a possible differential evolution of bright vs faint galaxies \citep[see also][]{Oesch16}.

At $z\sim10$, we are now in a situation that is similar to where we were at $z\sim7$ before the advent of WFC3/IR. Given the immediate revolution within the first few weeks of WFC3/IR data, it can be expected that $JWST$ will provide a similar jump in our knowledge at $z\gtrsim10$ within the first few weeks of operation. However, it is nevertheless important to realize that the low number density of $z\sim10$ sources found with $HST$ and the very rapid evolution of the UV LF seen at $z>8$ will imply that $JWST$ imaging surveys have to be planned carefully in order to be able to push our observational frontier of galaxies to well beyond 400 Myr after the Big Bang.

\vspace*{0.5cm}

\acknowledgments{ 
The authors thank the referee for a very constructive report, which helped to improve the manuscript.
The authors further thank P. Dayal, S. Wilkins, W. Cowley, S. Mutch, Ch. Liu, P. Behroozi, P. Laursen, and D. Ceverino for providing their latest model results. We also thank P. van Dokkum and M. Franx for illuminating and constructive discussions related to this work. 
PO acknowledges support from the Swiss National Science Foundation through the SNSF Professorship grant 157567 ‘Galaxy Build-up at Cosmic Dawn’.
We are grateful to the directors of STScI and SSC to execute the dedicated Frontier Field program. 
This work utilizes gravitational lensing models produced by PIs Bradac, Ebeling, Merten \& Zitrin, Sharon, and Williams funded as part of the HST Frontier Fields program conducted by STScI. STScI is operated by the Association of Universities for Research in Astronomy, Inc. under NASA contract NAS 5-26555. The lens models and datasets were obtained from the Mikulski Archive for Space Telescopes (MAST).}

Facilities: \facility{HST (ACS, WFC3), Spitzer (IRAC)}.

%\clearpage

\appendix

\section{Additional, Plausible HFF Candidates}

In Table \ref{tab:phot_extras}, we list a few additional candidate $z\sim10$ sources that satisfy our color criteria, but that were nevertheless excluded in our final analysis due to insufficient S/N and other criteria that put their reality into question. For all of these sources, we have performed additional, manual S/N measurements in circular apertures on the ICL subtracted images using the \texttt{iraf} task \texttt{qphot}, which confirmed that the S/N was indeed less than threshold of 6. Several of these sources additionally show extended morphologies (particularly M0416-874), which appear to be affected by correlated background noise on scales of 0\farcs5, further questioning their reliability.

\setlength{\tabcolsep}{0.04in} 
\begin{deluxetable*}{lccccccc}
\tablecaption{Photometry of Plausible, Additional $z\sim9-10$ Candidates in the HFF Data \label{tab:phot_extras}}
\tablewidth{0.8\linewidth}
\tablecolumns{7}
%\tabletypesize{\tiny}

\tablehead{\colhead{ID} & R.A. & Decl. &\colhead{$H_{160}$}  & \colhead{$J_{125}-H_{160}$}  & \colhead{S/N$_{160}$}    &  \colhead{Reason for Exclusion}}

\startdata

\cutinhead{\textit{MACS0416}}
M0416-874  &  04:16:10.43  &  -24:05:02.0  &  $27.76\pm0.20$ & $1.5 \pm 0.5 $  & 5.5 & low S/N \\

\cutinhead{\textit{MACS1149}}

M1149-3689 & 11:49:35.54 & 22:24:46.3 & $28.27\pm0.31$ & $> 1.7$ & 5.7 & low S/N \\ % 3689
M1149par-2407  &  11:49:44.71  &  22:18:36.7  &  $28.47\pm0.34$ & $> 1.7$  & 5.1  & low S/N  \\ 
M1149par-2491  &  11:49:43.70  &  22:18:40.3  &  $28.54\pm0.18$ & $> 1.7$  & 5.6  & low S/N \\

\cutinhead{\textit{Abell S1063}}
As1063par-2572  &  22:49:20.05  &  -44:31:55.8  &  $28.83\pm0.31$ & $> 1.5$   & 5.2 & low S/N

\enddata

\end{deluxetable*}

\bibliography{MasterBiblio}
\bibliographystyle{apj}

\end{document}